# Diagnostics of the condensate fraction in a clustered supersonic argon jet

Yu. S. Doronin, A. A. Tkachenko, V. L. Vakula, G. V. Kamarchuk

B. Verkin Institute for Low-Temperature Physics and Engineering

of the National Academy of Sciences of Ukraine,

Kharkiv 61103, Ukraine

E-mail: doronin@ilt.kharkov.ua

## Abstract

A new method for determining the condensate fraction and cluster density in absolute units has been proposed and tested for a supersonic argon jet, which can also be applied to supersonic jets of other gases. The method is based on measuring the absolute intensities of the Ar II resonance lines (93.2 and 92.0 nm) when the supersonic jet is excited by an electron beam with constant current density. Knowing the absolute intensities of the lines and their emission cross sections, we determined the density of the noncondensed atomic component in the supersonic argon jet and the evolution of the condensate fraction over the whole temperature range investigated.

## 1. Introduction

Supersonic cluster flows of rare gases have been actively used in many fields of science and technology for several decades [1-4]. To fully exploit the potential of nanoclusters in science and technology, it is essential to understand and control how their structure, composition, and physical and chemical properties change with size. Since nanoclusters can range from small groups of just a few atoms to larger nanocrystals containing many particles, their study is a non-trivial task. It requires sophisticated equipment and the latest experimental techniques [5].

Despite many fundamental studies of the condensation process in supersonic jets, the determination of the main parameters of the cluster beam, such as the mass fraction of condensate, cluster density, average cluster size, and the ratio of monomers and clusters in the jet, remains a difficult task. These parameters can be determined by combining several optical diagnostic techniques. For example, Rayleigh scattering, based on elastic scattering of light by particles, provides information on the size and density of gas clusters [6]. Mie scattering offers an estimate of the size distribution of clusters [7], and laser-induced fluorescence (LIF) helps measure a supersonic gas jet's density and flow pattern [8]. Additionally, these methods use rather complex, highly sensitive equipment and require time-consuming analytical modelling and mathematical calculations.

## 2. Experiment

An electron-excited supersonic jet of argon gas is a compact, intense source of vacuum ultraviolet radiation (gas-jet source, or GJS). With the atomic composition of the jet, VUV radiation is mainly generated by resonant transitions between the ground and excited states of atoms and ions. In jets containing clusters, the dominant contribution to the VUV flux is given by continua emitted by argon excimers. Figure 1 shows the emission spectrum of a supersonic argon jet in the wavelength range of 90-150 nm at gas pressure at the nozzle inlet $P_0 = 0.2$ MPa and temperature $T_0 = 300$ K. An electron beam with an energy of 1 keV and a current of 20 mA crossed the supersonic jet at a distance of 5 mm from the nozzle exit section.

The spectrum contains resonance lines emitted by argon ions Ar II ($\lambda$=92.0 nm, $\lambda$=93.2 nm) and atoms Ar I ($\lambda$=104.8 nm, $\lambda$=106.7 nm), as well as cluster continua with maxima $\lambda$=107. 5, 109 nm (1 and 2) emitted by excimers from partially vibrationally relaxed states and a continuum with maximum $\lambda$=127 nm (M) emitted by excimers in the vibrationally relaxed state [9]. Since the lines have much higher intensities than the continuum region under the above conditions, they are multiplied by the coefficients given on the spectrum for ease of presentation. It should be noted that the continuum centred at $\lambda$=127 nm is present in the emission spectra of most plasma VUV sources operating at sufficiently high argon pressures [10] and in the luminescence spectra of liquid and solid argon [11, 12].

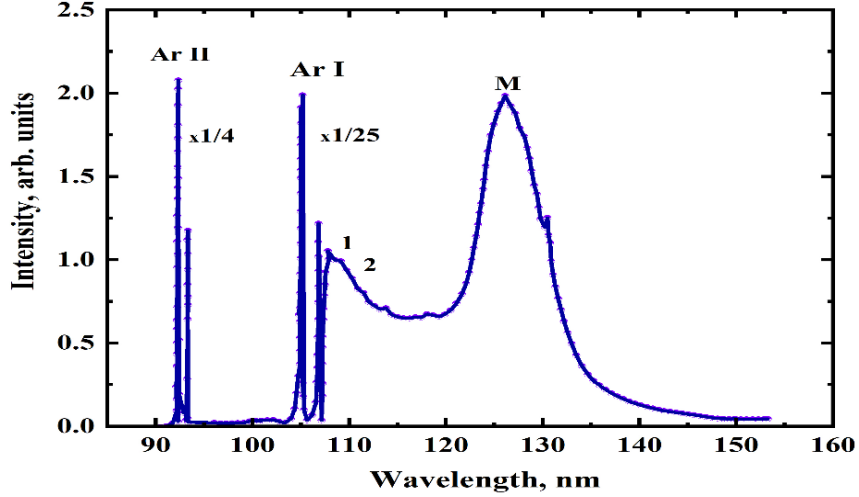

*Fig. 1.* Emission spectrum of a supersonic argon jet excited by an electron beam with energy 1 keV at gas pressure $P_0$ = 0.2 MPa and temperature $T_0$ = 300 K at the nozzle inlet.

Absolute measurements of the radiation flux, carried out using the methodology [13] for this mode of the GJS operation, allow us to use GJS as a calibrated source in the VUV region of the spectrum. The integral radiation flux $Q$ (photon/s·sr) of a supersonic argon jet at given parameters of gas flow through the nozzle is measured in absolute units in the wavelength range of 50-200 nm by a silicon detector SXUV-100 with known spectral sensitivity. Simultaneously, the relative intensity distribution $I(\lambda)$ in the argon jet emission spectrum is recorded by a vacuum monochromator in the same spectral range of 50-200 nm. The obtained spectrum is corrected for the efficiency of the vacuum monochromator. After determining $Q$ and $I(\lambda)$, the spectral distribution of the radiation flux density $Q(\lambda)$ of the supersonic argon jet in the whole investigated wavelength range is found:

$$Q(\lambda) = \frac{Q \cdot I(\lambda)}{\int_{50}^{200} I(\lambda) d\lambda}, \qquad (1)$$

where $\int_{50}^{200} I(\lambda) d\lambda$ is the area occupied by the spectrum in the 50-200 nm region.

In accordance with the above methodology, the intensity of the Ar II resonance lines (92.0 and 93.2 nm) was measured when the supersonic jet was excited with an electron beam ($I_{el}$ = 20 mA, $E_{el}$ = 1 keV) at distances of 5 and 30 mm from the nozzle exit. Figure 2 shows the spectrum in the 91-94 nm wavelength range recorded at a resolution of 1 Å for a distance of 30 mm. The argon pressure and gas temperature at the nozzle inlet were $P_0 = 0.1$ MPa and $T_0=400$ K, corresponding to the jet's atomic composition. The radiation flux $\Phi$ ($\lambda$=92.0 nm) was 3.2 $10^{11}$ photons/s, and $\Phi$ ($\lambda$=93.2nm) was 1.7 $10^{11}$ photons/s.

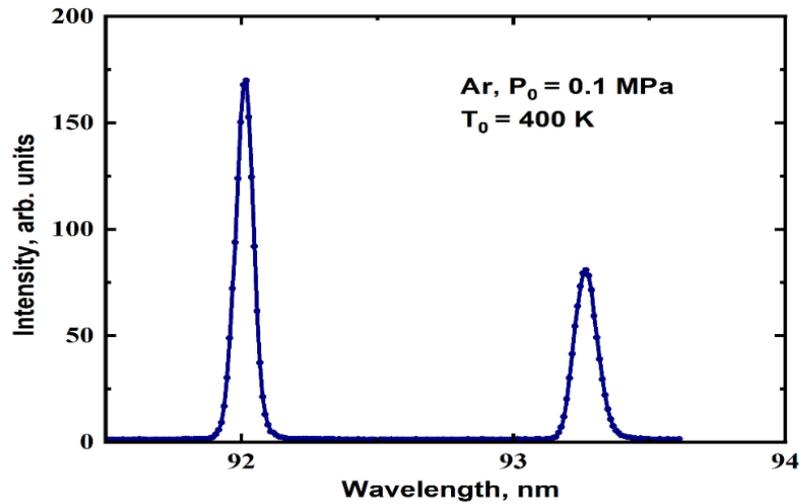

***Fig 2.*** *Emission spectrum of supersonic argon jet excited by electron beam with energy 1 keV at gas pressure $P_0$=0.1 MPa and temperature $T_0$=400 K at the nozzle inlet.*

At the same distance at pressure $P_0 = 0.1$ MPa, the dependence of the Ar II (92.0 nm) line intensity on the temperature $T_0$ in the interval of 150-500 K was obtained, which is shown in relative units in Fig. 3. It should be noted that a similar dependence for Ar II (93.2 nm) has the same form.

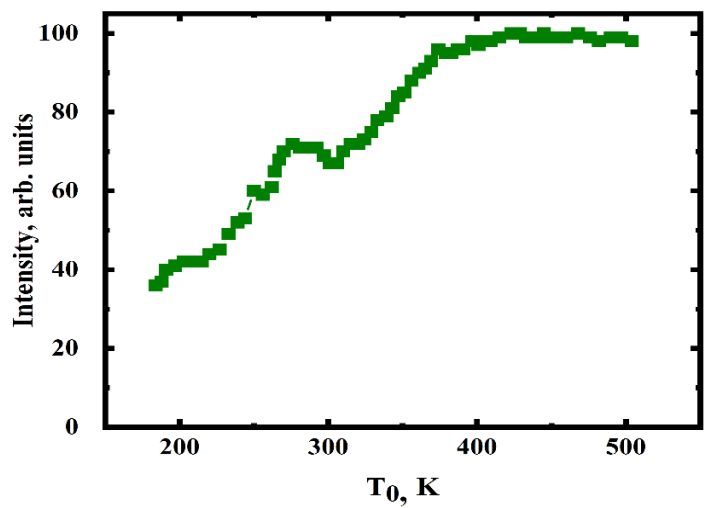

***Fig. 3.*** *Intensity of the Ar II line (92.0 nm) as function of temperature $T_0$ at pressure $P_0$=0.1 MPa.*

Under the same conditions and in the same temperature range, the intensity dependence of the cluster continuum with the maximum at λ = 127 nm (see Fig. 4) corresponding to transitions from the excited states of the $Ar_2^*$ molecule ($^{1,3}\Sigma_u^+$) to the repulsive branch of the ground state $^1\Sigma_g$ of the $Ar_2$ [9] molecule was measured.

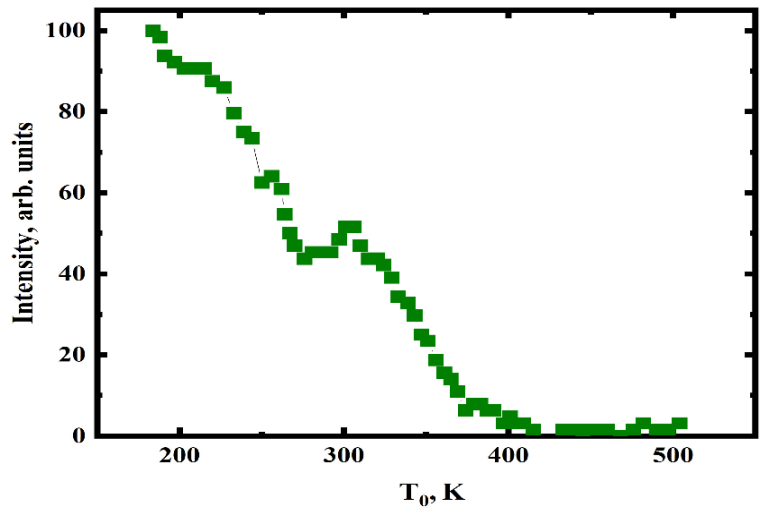

***Fig. 4.*** *Intensity of the cluster continuum at λ=127 nm versus temperature $T_0$ at pressure $P_0$=0.1 MPa.*

Our analysis has shown that the general form of the dependence of the intensity of the Ar II resonance lines (92.0 and 93.2 nm) on temperature recorded at a constant pressure of $P_0$ = 0.1 MPa at different distances from the nozzle exit remains constant. The ratio of their intensities also does not change. It should be emphasised that the canonical correlation analysis performed using the STATGRAPHICS plus software package revealed a strong inverse correlation ($r \approx -1$) between the obtained temperature dependences for the line intensities and the cluster continua (see Figs. 3, 4). This indicates the presence of a strong correlation between the intensities of the ion line and cluster continuum: when the intensity of the cluster continuum increases with decreasing temperature (with increasing cluster size), a natural decrease in the ion line intensity is observed. Thus, the revealed inverse correlation ($r \approx -1$) is a strong argument in favour of the extracluster nature of ion emission.

The probability of electron-induced emission for the case of excitation of the argon supersonic jet by electrons with a fixed energy can be described by the photoemission cross section determined by the following equation:

$$\sigma(\lambda) = \frac{1}{n\, l\, \frac{i}{e}} \frac{4\pi}{\Omega} \Phi(\lambda), \qquad (2)$$

where $\sigma(\lambda)$ is the absolute photoemission cross section [cm$^2$], $n$ is the atomic jet density [cm$^{-3}$]; $l$ is the electron beam length in the area of jet excitation [cm]; $\Omega$ is the solid angle of observation [sr]; $I$ is the electron beam current [A]; $e$ is the electron charge [C], $\Phi(\lambda)$ is the absolute radiation flux for the emission under study.

Using the emission cross section $\sigma(\lambda)$ for the $\lambda$ = 92.0 nm line measured for excitation by 1-keV electrons in Refs. [14,15] and our obtained emission flux $\Phi$ ($\lambda$ = 92.0 nm) = 3.2 10$^{11}$ photons/s, we estimated the concentration of argon atoms $n$ in the investigated region of a jet characterized by a purely atomic composition ($P_0$ = 0.1 MPa and $T_0$ = 400 K):

$$n = \frac{1}{\sigma(\lambda)l\frac{i}{e}}\frac{4\pi}{\Omega}\Phi(\lambda), \qquad (3)$$

It should be noted that the contribution of cascade processes, self-absorption [14], and desorption to the intensity of the $\lambda$ = 92.0 nm line does not exceed 10% and, therefore, does not significantly affect the accuracy of the determination of the concentration of atoms $n$ in the investigated region of the jet. The analysis of the processes of desorption of atoms from clusters and their contribution to the line intensity will be discussed in detail in the next publication.

Similarly, we can determine $n$ at any temperature in the range of 150-400 K using the dependence of the intensity of the $\lambda$ = 92.0 nm line on temperature (Fig. 3). Then, knowing the density of atoms in the jet in the absence of clustering, which on the curve for the $\lambda$ = 92 nm line corresponds to temperatures of around 400 K, and its change with decreasing temperature, we can determine the condensate fraction - the ratio of the number of atoms participating in the formation of clusters to the number of noncondensed atoms in the temperature range of 150-400 K. The dependence of the condensate fraction on temperature is shown in Fig. 5.

Having estimated the number of atoms participating in clustering and the average cluster sizes experimentally determined for similar jet parameters [16,17], we calculated the density of clusters in the supersonic jet Ar as a function of their size (see Fig. 6). The curve shows a non-monotonic dependence of density of clusters on their average size, which is characteristic of the systems in which cluster formation transitions from condensation processes dominated by nucleation to those dominated by coalescence. At the initial stage of the clustering process, the density of clusters increases rapidly as the number of condensation nuclei increases and reaches a maximum for an average cluster size of $N_{cl}\approx$150 at/cl.

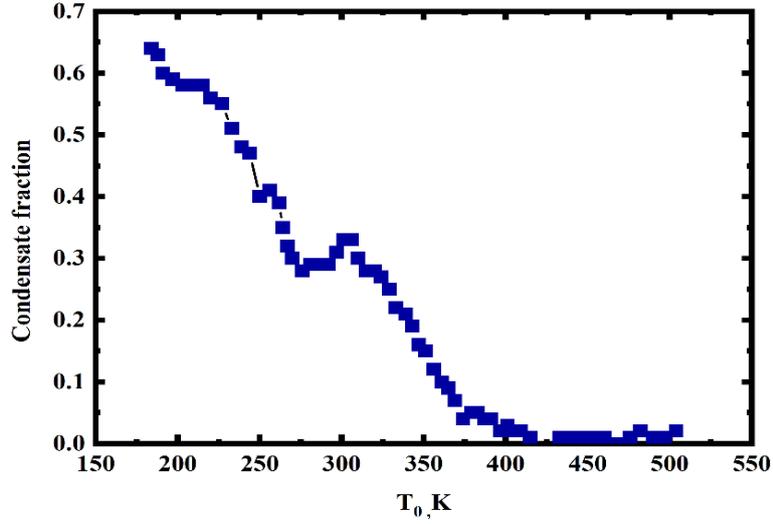

*Fig. 5.* Condensate fraction versus gas temperature $T_0$ (electron beam with energy 1 keV and current 20 mA crossed the argon supersonic jet at 30 mm from the nozzle exit, pressure at the nozzle inlet being $P_0 = 0.1$ MPa).

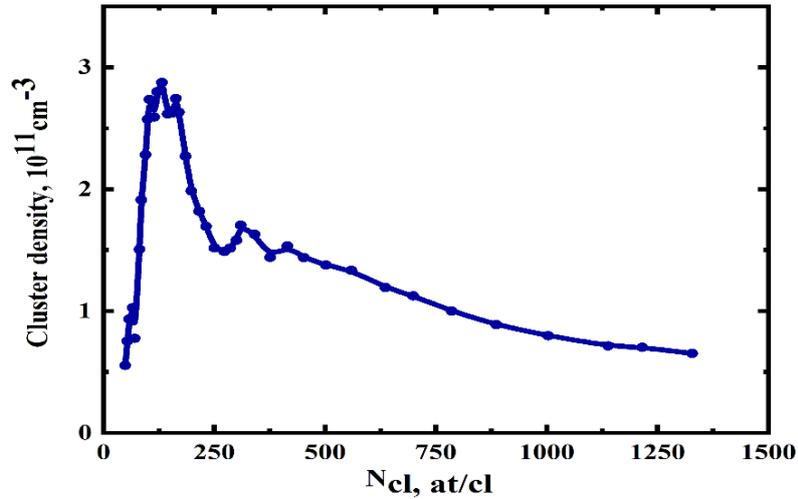

*Fig. 6.* Density of Ar clusters in the investigated region of the supersonic gas jet.

The observed secondary peaks can result from fluctuations of local supersaturation or be caused by fragmentation of small unstable clusters. Starting from $N_{cl}\approx 400$ at/cl, the coalescence process becomes predominant, increasing the average size of clusters and decreasing their number. Thus, the obtained dependence of cluster density on their size reflects the dynamics of the cluster formation and growth in the cluster size interval of 10-1400 at/cl.

## 3. Conclusion

This paper presents a novel method for determining an argon supersonic jet's condensate fraction and cluster density. The technique, which utilizes the absolute intensities of the Ar II resonance lines, provides a reliable basis for measuring the density of noncondensed (atomic) components and allows one to control the condensate fraction in different temperature ranges, which is crucial for characterising the behaviour of jets under various conditions. The method's principles make it versatile enough for other gases [18], improving understanding of condensation processes in broader applications, including aerospace engineering and environmental science.

Our studies have shown that the operation of the gas-jet source is stable. The total instability of the source parameters with time does not exceed 3%. To calibrate the GJS, we used a silicon photodiode SXUV-100, whose spectral sensitivity error in the wavelength range of 50 – 200 nm does not exceed 10 %. As a result, the relative error in determining the investigated parameters of supersonic flow using the proposed technique was about 20 %. The error can be reduced by a more accurate calibration of the spectrometer-detector system.